\documentclass[aps,prb,twocolumn,groupedaddress]{revtex4}

\usepackage{graphicx}% Include figure files
\usepackage{dcolumn}% Align table columns on decimal point
\usepackage{bm}% bold math

\begin{document}

% Use the \preprint command to place your local institutional report
% number in the upper righthand corner of the title page in preprint mode.
% Multiple \preprint commands are allowed.
% Use the 'preprintnumbers' class option to override journal defaults
% to display numbers if necessary
%\preprint{}

%Title of paper
\title{Soft X-ray magnetic circular dichroism study of the ferromagnetic Cr$_{1-\delta}$Te}

% repeat the \author .. \affiliation  etc. as needed
% \email, \thanks, \homepage, \altaffiliation all apply to the current
% author. Explanatory text should go in the []'s, actual e-mail
% address or url should go in the {}'s for \email and \homepage.
% Please use the appropriate macro foreach each type of information

% \affiliation command applies to all authors since the last
% \affiliation command. The \affiliation command should follow the
% other information
% \affiliation can be followed by \email, \homepage, \thanks as well.
\author{Koichiro Yaji}
%\altaffiliation[Also at ]{Department of Physical Sciences,Graduate School of Science, Hiroshima University}%Lines break automatically or can be forced with \\
\author{Akio Kimura}%
\email{akiok@hiroshima-u.ac.jp}
\author{Chiyuki Hirai}
\author{Hitoshi Sato}
\author{Masaki Taniguchi}
\affiliation{%
Graduate School of Science, Hiroshima University, 1-3-1 Kagamiyama, Higashi-Hiroshima 739-8526, Japan\\
}%

\author{Michie Koyama}
\affiliation{
Kure National College of Technology, Agaminami 2-2-11, Kure 737, Japan\\
}%

\author{Kenya Shimada}
\affiliation{
Hiroshima Synchrotron Radiation Center,Hiroshima University, 2-313 Kagamiyama, Higashi-Hiroshima 739-8526, Japan \\
}%

\author{Arata Tanaka}
\affiliation{
Department of Quantum Matter, ADSM, Hiroshima University, 1-3-1 Kagamiyama, Higashi-Hiroshima 739-8526, Japan \\
}%

\author{Takayuki Muro}
\affiliation{
Japan Synchrotron Radiation Research Institute, Mikazuki, Hyogo 679-5143, Japan \\
}%

\author{Shin Imada, Shigemasa Suga}
\affiliation{
Graduate School of Engineering Science, Osaka University, 1-3 Machikaneyama, Toyonaka, Osaka 560-8531, Japan \\
}%

%Collaboration name if desired (requires use of superscriptaddress
%option in \documentclass). \noaffiliation is required (may also be
%used with the \author command).
%\collaboration can be followed by \email, \homepage, \thanks as well.
%\collaboration{}
%\noaffiliation

\date{\today}

\begin{abstract}
The $2p$ core excited XAS and XMCD spectra of Cr$_{1-\delta}$Te with several concentrations of $\delta$=0.11-0.33 have been measured.
The observed XMCD lineshapes are found to very weakly depend on $\delta$.
The experimental results are analyzed in terms of the configuration-interaction picture with consideration of hybridization and electron correlation effects.
The calculated result shows that CrTe can be classified into a charge transfer type material and created holes preferably exist in Te $5p$ orbitals in Cr deficient materials Cr$_{1-\delta}$Te, which are in consistence with the observed XMCD feature and the reported band structure calculation.
\end{abstract}

% insert suggested PACS numbers in braces on next line
% insert suggested keywords - APS authors don't need to do this
%\keywords{}

%\maketitle must follow title, authors, abstract, \pacs, and \keywords
\maketitle

% body of paper here - Use proper section commands
% References should be done using the \cite, \ref, and \label commands
\section{Introduction}
% Put \label in argument of \section for cross-referencing
%\section{\label{sec:level1}}
%\subsection{}
%\subsubsection{}
Chromium chalcogenides Cr$_{1-\delta}$X (X=S, Se, Te) with metal-deficient NiAs type crystal structures show various magnetic and electronic properties.\cite{LB}
Among them, chromium tellurides Cr$_{1-\delta}$Te are ferromagnets with Curie temperatures of 170-360K.
Ipser et al. systematically studied the phase diagram of Cr-Te system and determined the relationship between the phase and crystal structures.\cite{Ipser83}
The Cr$_{1-\delta}$Te with $\delta<0.1$ form the hexagonal NiAs crystal structure, while the Cr$_{3}$Te$_{4}$($\delta$=0.25) and Cr$_{2}$Te$_{3}$($\delta$=0.33) form the monoclinic and the trigonal crystal structures, where Cr vacancies occupy in every second metal layer.
The effective paramagnetic moments described as $\mu_{\rm eff}= g\sqrt{J(J+1)}$ are $\sim 4-4.9\mu_{\rm B}$ for the hexagonal Cr$_{1-\delta}$Te, $\sim$4.2$\mu_{\rm B}$ for Cr$_{3}$Te$_{4}$ and $\sim$3.96$\mu_{\rm B}$ for Cr$_{2}$Te$_{3}$, which have been explained by the spin moments $2\sqrt{S(S+1)}$ only with $S=2$ for $\delta\sim$ 0 and with $S= 3/2$ for $\delta$=0.333.\cite{Lotgering57, Hirone60, Ohsawa72, Grazhdankina70, Hashimoto69, Yamaguchi72, Andresen63, Andresen70, Hashimoto71, Kanomata00}
However, the ordered magnetic moments evaluated from the magnetization measurements show much smaller values such as 2.4-2.7$\mu_{\rm B}$ for the hexagonal Cr$_{1-\delta}$Te, 2.35$\mu_{\rm B}$ for Cr$_{3}$Te$_{4}$($\delta$=0.25) and 2.0$\mu_{\rm B}$ for Cr$_{2}$Te$_{3}$($\delta$=0.333) than those expected from the ionic model.\cite{Lotgering57, Hirone60, Ohsawa72, Grazhdankina70, Hashimoto69, Yamaguchi72, Andresen63, Andresen70, Hashimoto71, Kanomata00, Ohta93, Kanomata98}
According to neutron-diffraction studies, the small value of saturation magnetization is partly explained if we take the spin canting into account for $\delta=0.125, 0.167$ and $0.25$.\cite{Andresen70}
The magnetic moment induced on the Cr ion for $\delta$=0.25 is close to an integral number of Bohr magnetons, suggesting the existence of mixed valence Cr. \cite{Andresen70}
However for Cr$_{2}$Te$_{3}$($\delta$=0.33), the ordered magnetic moment of $2.65-2.70\mu_{B}$, deduced from the neutron diffraction, is much smaller than that calculated using the ionic model, $3\mu_{B}$, suggesting the itinerant nature of the $d$ electrons.\cite{Andresen70, Hamasaki75}

The experimental Hall resistivity of Cr$_{0.9}$Te as a function of applied magnetic field up to 3T in the temperature range of 4.2-340K can be fitted with the ordinary formula containing the normal and anomalous Hall coefficients as for the most ferromagnetic materials.\cite{Dijikstra89}
The temperature dependent anomalous Hall coefficient $R_{\rm S}$ can be understood for the collinear ferromagnets.\cite{Dijikstra89}
The number of carriers, which has been estimated from the normal Hall coefficient $R_{0}$ is +0.9/formula unit, suggesting hole carriers in Cr$_{0.9}$Te.\cite{Dijikstra89}
In contrast, the Hall resistivity of Cr$_{3}$Te$_{4}$ shows a different behavior from that of Cr$_{0.9}$Te below $T$=100K, which suggests that there exists a complex magnetic structure in the monoclinic Cr$_{3}$Te$_{4}$ below $T$=100K.\cite{Oda01}

Remarkable magnetovolume effect of Cr$_{1-\delta}$Te has been reported, where the spontaneous magnetization decreases rapidly with increasing pressure as reported for Cr$_{2}$Te$_{3}$ and Cr$_{0.92}$Te.\cite{Kanomata98, Kanomata00}
It has also been reported very recently that the ferromagnetism disappears at 7GPa, and under pressure above 13GPa the structural phase transition occurs from the NiAs type structure to the orthorhombic MnP type structure.\cite{Ishizuka01, Eto01}

Electronic specific heat coefficients $\gamma$ are also much dependent on $\delta$.
The estimated $\gamma$ for Cr$_{5}$Te$_{6}$($\delta$=0.167), Cr$_{3}$Te$_{4}$($\delta$=0.25) and Cr$_{2}$Te$_{3}$($\delta$=0.333) are 10, 1 and 4 mJ/atom/K$^{2}$.\cite{Gronvold64}
The $\gamma$ for Cr$_{3}$Te$_{4}$ is quite close to the predicted value 1.0$\sim$1.4 mJ/atom/K$^{2}$ using the calculated density of states (DOS) at the Fermi level ($E_{\rm F}$), while that for Cr$_{2}$Te$_{3}$ is much larger than the predicted one, $\gamma\sim0.82-0.96$ mJ/atom/K$^{2}$.\cite{Dijikstra89, Ishida}
Such large $\gamma$ values suggest that electron correlation effects are also important in Cr$_{5}$Te$_{6}$ and Cr$_{2}$Te$_{3}$.
The electron correlation effects in these "itinerant ferromagnets" Cr$_{1-\delta}$Te has been discussed with the photoemission spectra.\cite{Shimada96}
It has been pointed out that the spectral weight in $E_{\rm B}$=2-4eV observed in the photoemission spectra of Cr$_{0.95}$Te and Cr$_{3}$Te$_{4}$ can not be explained by the theoretical band structure calculation, and the intensity at $E_{\rm F}$ is found to be smaller than the theoretical DOS.
On the other hand, the spectral weight in $E_{\rm B}$=2-4eV has been reproduced with the configuration interaction cluster-model calculation, indicating the importance of the electron correlation effect in Cr$_{1-\delta}$Te.\cite{Shimada96}

It is known that soft X-ray core absorption spectroscopy (core XAS) is a powerful tool to study the element specific valence electronic states of materials.
X-ray magnetic circular dichroism (XMCD) in the core XAS spectrum provides us with useful information on the element specific spin and orbital magnetic moments with use of the "sum rule".\cite{Thole92, Carra93, Chen95}
%%%%%%%%%%%%%%Moved from SecIV and Revised at 2002/12/30 (below)%%%%%%%%%%%%%%%
It is widely known that the spectral shapes of the XAS and XMCD spectra are strongly dependent on the electronic states or the electronic configuration of the derived atom.
Besides, these lineshapes can be remarkably affected by the intersite hybridization between the surrounding atoms or by the band structure of the crystal.
In other words, the spectra can be a good fingerprint of the electronic states inside the crystals.
For example, the valency (or the electron number) can be determined by the XAS and XMCD spectral lineshapes.
%%%%%%%%%%%%%% Moved from SecIV and Revised at 2002/12/30 (above)%%%%%%%%%%%%%%%
One can also determine several physical parameters such as the Coulomb repulsion energy, the charge transfer energy as well as the hybridization energy from the analyses of experimental XAS and XMCD spectral lineshapes.
In order to understand how the electronic states are related to the $\delta$ dependence of the magnetic and electronic properties, we have done Cr $2p$ XAS and XMCD study for Cr$_{8}$Te$_{9}$($\delta=0.11$), Cr$_{5}$Te$_{6}$($\delta=0.17$), Cr$_{3}$Te$_{4}$($\delta=0.25$) and Cr$_{2}$Te$_{3}$($\delta=0.33$).

\section{Experimental}
Metallographical studies for the Cr-Te system have shown that the existence of stoichiometric 'CrTe' is ruled out.\cite{Ipser83}
For example, the stability range of the phase with NiAs-type structure extends from 52.4 to 53.3 at.\% Te.
Polycrystalline samples were synthesized from mixed powders of constituent elements.
They were sealed in evacuated silica tubes, which were heated for a week at 1000$^{\circ}C$.
After this, the samples were ground and sealed in silica tubes again and heated for 2 hours at 1450$^{\circ}$C and then cooled gradually to 1000$^{\circ}$C and finally quenched into water.\cite{Koyama00}
The stoichiometry and the homogeneity of Cr$_{1-\delta}$Te have been estimated by means of Electron Probe Micro-Analysis (EPMA).
X-ray diffraction studies confirmed that all of the samples were in a single phase.

Cr $2p$ core absorption spectroscopy (XAS) and X-ray magnetic circular dichroism (XMCD) spectra were measured at BL25SU of SPring-8 in Japan.\cite{Suga01, Suga02, Saitoh01, Saitoh02}
Circularly polarized light was supplied from a twin-helical undulator, with which almost 100\% polarization was obtained at the peak of the first-harmonic radiation.
After having set the two undulators to opposite helicity, helicity reversal was realized by closing one undulator and fully opening the other.\cite{Saitoh01}
Cr $2p$ XAS spectra were measured by means of the total photoelectron yield method by directly detecting the sample current while changing the photon energy $h\nu$.
The photon energy resolution was set to $E/\Delta E=5000$ for the Cr $2p$ core excitation regions.
The measurement was performed in the Faraday geometry with both the incident light and the magnetization perpendicular to the sample surface.
We used two pairs of permanent dipole magnets with holes for passing the excitation light.
The external magnetic field of $\sim1.4T$ at the sample position was alternatingly applied by setting one of the two dipole magnets on the optical axis by means of a moter-driven linear feedthrough.
The XMCD spectra were taken for a fixed helicity of light by reversing the applied magnetic field at each $h\nu$.
In the present paper, the XMCD spectrum is defined as $I_{+}-I_{-}$, where $I_{+}$ and $I_{-}$ represent the absorption spectra for the direction of magnitization (which is opposite to the direction of the majority spin) parallel and antiparallel to the photon helicity, respectively.\cite{Suga01}
Clean surfaces were obtained by {\it in situ} scraping of the samples with a diamond file under ultra high vacuum condition (3$\times$10$^{-8}$ Pa).
The cleanliness of the sample surfaces was first checked by the disappearance of a typical structure related to Cr oxides.
We could also check the degree of contamination from the magnitude of the XMCD signal, because its amplitude grew and finally saturated when the sample surface became clean enough.
We considered that the unscraped or contaminated surface was covered with antiferromagnetic or paramagnetic compounds such as Cr$_{2}$O$_{3}$, which hardly contribute to the XMCD spectrum.
It is generally known that the the total photoelectron yield reflects the absorption spectrum in the core-excitation region.
The temperatures during the measurement were $\sim$110K for all of the samples.

\section{CI Cluster Model Calculation with Full Multiplets}
The CI cluster model calculation has been done with a program code developed by A. Tanaka by means of the recursion method.
The detailed procedures are described elsewhere.\cite{Tanaka94}
Slater integrals have been calculated by Cowan's code and the calculated values are listed in Table \ref{SI}.
In the calculation, the Slater integrals are scaled down to 80\% of the listed values to take into account intra-atomic relaxation effect.

\section{Results and Discussion}
%*************************************
\begin{figure} 
\includegraphics{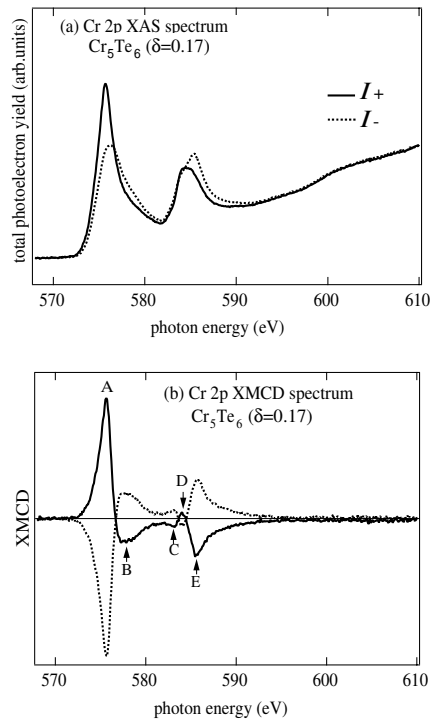}%
\caption{(a) Cr $2p$ XAS spectra ($I_{+}$ and $I_{-}$) with different helicities of incident radiation.
(b)XMCD spectrum ($I_{+}-I_{-}$) of Cr$_{5}$Te$_{6}$ ({\it solid line}).
The XMCD spectrum taken by reversing the helicity of the incident radiation is also shown ({\it dashed line}).
}
\end{figure}
%*************************************
The XAS and XMCD spectra in the Cr $2p$ core excitation region have been measured for Cr$_{8}$Te$_{9}$ ($\delta=0.11$), Cr$_{5}$Te$_{6}$($\delta=0.17$), Cr$_{3}$Te$_{4}$($\delta=0.25$) and Cr$_{2}$Te$_{3}$($\delta=0.33$).
The Cr $2p$ XAS ($I_{+}$ and $I_{-}$) spectra with both helicities of incident radiation and the XMCD ($I_{+}-I_{-}$) spectra of Cr$_{5}$Te$_{6}$ are shown in Fig.1 (a) and (b).
It is found that the $2p_{3/2}$ and the $2p_{1/2}$ core absorption peaks are located at about 576 and 585 eV, and the broad hump is found around the photon energy of 600-605 eV.
Since the core level binding energies ($E_{\rm B}$) of the Cr $2p$ and Te $3d$ levels are quite close to each other, one may expect the overlapping of these absorption edges.\cite{Fuggle80}
%%%%%%%%%%%%%%Revised at 2002/12/30 (below)%%%%%%%%%%%%%%%
We expect much lower Te $3d \rightarrow 5p$ absorption cross section ($<$5\%) compared to that of the Mn $2p \rightarrow 3d$ absorption as observed, i.e., in Te $4d \rightarrow 5p$ absorption spectrum of MnTe$_{2}$.\cite{Kaznacheyev98} 
%%%%%%%%%%%%%%Revised at 2002/12/30 (above)%%%%%%%%%%%%%%%
We also expect that the observed Cr $2p$ XAS fine structures are almost unaffected by the Te $3d$ XAS spectrum because the Te $3d$ core absorption is expected to be very broad due to the wide conduction band derived from the strong Cr $3d$-Te $5p$ hybridization.
Therefore one can assume that the Te $3d$ core absorption spectrum behaves like a background for the Cr $2p$ core absorption structures and the observed XAS spectra in the present photon energy range mostly reflect the Cr $2p$ core absorption.
However, the broad hump at 600-605 eV can be still assigned to the Te $3d\rightarrow5p$ absorption mainly because the observed XMCD asymmetry is negligible in this energy region.
It is found that the intensity of $I_{+}$ is larger than that of $I_{-}$ in the $2p_{3/2}$ core absorption region, whereas the intensity of $I_{+}$ is smaller than that of $I_{-}$ in the $2p_{1/2}$ region.
Besides, the spectral weight of $I_{+}$ is shifted to lower energy compared to $I_{-}$ in both $2p_{3/2}$ and $2p_{1/2}$ regions.
This derives the complicated XMCD ($I_{+}-I_{-}$) structures as shown by the solid line in Fig.1 (b).
Here, the spectrum ({\it solid line}) shows remarkable XMCD with positive sign at the $2p_{3/2}$ core absorption edge ($h\nu$=575.5eV) as marked with A, which is followed by the smaller asymmetry with negative sign ($h\nu$=578eV) as represented by B.
It is noticed that the XMCD signal does not reach zero even in the region between the spin-orbit split $2p$ components.
There is still finite and negative XMCD on the lower energy side of the $2p_{1/2}$ absorption edge ($h\nu$=583eV).
Then one finds a small positive ($h\nu$=584eV) and a large negative ($h\nu$=585.5eV) XMCD peaks with increasing $h\nu$ as marked with C, D and E in Fig.1 (b).
To eliminate the possible instrumental asymmetries, we have taken the spectra by reversing the helicity of the incident radiation.
As a result of this procedure, it is found that the observed XMCD signals with opposite helicities of the incident lights are quite symmetric with respect to the zero line as shown by the solid  and dashed lines.
This means that the observed complicated structures of the XMCD spectrum are intrinsic signals.

In Fig.2, are shown the helicity averaged Cr $2p$ XAS spectra represented as $(I_{+}+I_{-})/2$ of (a)Cr$_{8}$Te$_{9}$($\delta$=0.11), (b)Cr$_{5}$Te$_{6}$($\delta$=0.17), (c)Cr$_{3}$Te$_{4}$($\delta$=0.25) and (d)Cr$_{2}$Te$_{3}$($\delta$=0.33).
In contrast to the spectrum of Cr$_{5}$Te$_{6}$ (Fig.2 (b)), one finds a shoulder on the higher $h\nu$ side of the $2p_{3/2}$ edge in the spectrum of Cr$_{8}$Te$_{9}$.
One also finds broad and small shoulder on the lower energy side of the $2p_{1/2}$ peaks in Cr$_{8}$Te$_{9}$.
In the cases of the higher $\delta$ compounds like Cr$_{5}$Te$_{6}$, Cr$_{3}$Te$_{4}$ and Cr$_{2}$Te$_{3}$, the line width of the $2p_{3/2}$ edge is narrower with some tail extending to the higher $h\nu$ compared to that of Cr$_{8}$Te$_{9}$.
Besides, the shoulder on the lower $h\nu$ side of the $2p_{1/2}$ edge has comparable spectral weight to that of the $2p_{1/2}$ main peak in the spectra of Cr$_{5}$Te$_{6}$, Cr$_{3}$Te$_{4}$ and Cr$_{2}$Te$_{3}$.
In addition, the broad humps are observed at the same $h\nu$ region near 600-605eV in all of the spectra as in the case of Cr$_{5}$Te$_{6}$.

%*************************************
\begin{figure}
\includegraphics{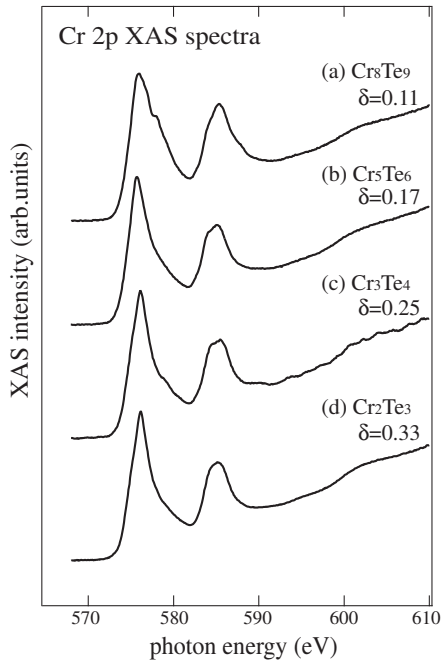}%
\caption{Helicity averaged Cr $2p$ XAS spectra $(I_{+}+I_{-})/2$ of (a)Cr$_{8}$Te$_{9}$, (b)Cr$_{5}$Te$_{6}$, (c)Cr$_{3}$Te$_{4}$ and (d)Cr$_{2}$Te$_{3}$.
}
\end{figure}
%*************************************

The XMCD spectra obtained as $I_{+}-I_{-}$ of (a)Cr$_{8}$Te$_{9}$, (b)Cr$_{5}$Te$_{6}$, (c)Cr$_{3}$Te$_{4}$ and (d)Cr$_{2}$Te$_{3}$ are shown in Fig.3.
We find that the overall lineshapes of the present XMCD spectra are similar to each other, in which all of the fundamental structures A-E are observed, but we observe small changes depending on $\delta$.
In the spectra of Cr$_{3}$Te$_{4}$ and Cr$_{2}$Te$_{3}$, the shoulder structures are present on the lower energy side of the Cr $2p_{3/2}$ absorption edge as denoted by F in Fig.3.
In addition, the negative structures on the higher energy side of the $2p_{3/2}$ main peak is sharper in the spectra of Cr$_{3}$Te$_{4}$ and Cr$_{2}$Te$_{3}$ than in those of Cr$_{8}$Te$_{9}$ and Cr$_{5}$Te$_{6}$.

%*************************************
\begin{figure}
\includegraphics{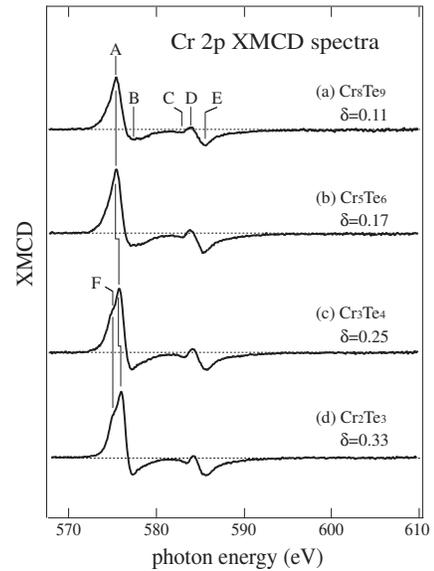}%
\caption{Cr $2p$ XMCD $(I_{+}-I_{-})$ spectra (a)Cr$_{8}$Te$_{9}$, (b)Cr$_{5}$Te$_{6}$, (c)Cr$_{3}$Te$_{4}$ and (d)Cr$_{2}$Te$_{3}$.
}
\end{figure}
%*************************************

First, we have tried to evaluate the contribution of the orbital magnetic moment $m_{\rm orb}$ with the use of the following sum rule\cite{Thole92, Chen95},
\begin{equation}
\displaystyle m_{\rm orb}=\frac{4}{3}\cdot\frac{\int_{L_{3}+L_{2}}(I_{+}-I_{-})dh\nu}{\int_{ L_{3}+L_{2}}(I_{+}+I_{-})dh\nu}\cdot(10-n_{d})
\end{equation}
, where $n_{d}$ represents $3d$ electron number.
The estimated $m_{\rm orb}$ value is turned out to be almost negligible, suggesting the quenched Cr $3d$ orbital magnetic moment in consistence with the measured $g$ value ($g\sim 2$).\cite{Grazhdankina70}

Next, in order to evaluate several physical parameters that control the physical properties of Cr$_{1-\delta}$Te, we have calculated the Cr $2p$ XAS and XMCD spectra by means of a configuration interaction (CI) cluster-model calculation with full multiplets assuming a [CrTe$_{6}$]$^{10-}$ cluster so as to reproduce the experimental spectra of Cr$_{1-\delta}$Te.
Here, the nominal $d$ electron numbers in Cr$_{8}$Te$_{9}$($\delta=$0.11), Cr$_{5}$Te$_{6}$($\delta=0.17$), Cr$_{3}$Te$_{4}$($\delta=0.25$) and Cr$_{2}$Te$_{3}$($\delta=0.33$) are $\sim$ 3.75, 3.60, 3.33 and 3.00 per Cr atom, respectively when we assume the Te valence to be $2-$.
Among them, the formal valency of Cr$_{8}$Te$_{9}$ is the closest to $2+$ ($3d^{4}$), which is the nominal valency of CrTe (mono telluride).
%%%%%%%%%%%%%%%%%%%%%%%%%%%%%%%%%%%%%%%%%%%%%%%%%%%%%%%%%%%%%%%
\begin{table*}%[H] add [H] placement to break table across pages
\caption{\label{SI}{\it Ab initio} Hartree-Fock values of the Slater integrals and spin-orbit coupling constants (in units of eV).In the actual calculation, the Slater integrals have been scaled to 80\% of these values to take into account intra-atomic relaxation effect.}
\begin{ruledtabular}
\begin{tabular}{lc|ccccccc}
&configuration&$F^{2}(d,d)$&$F^{4}(d,d)$&$F^{2}(p,d)$&$G^{1}(p,d)$& $G^{3}(p,d)$&$\xi(3d)$&$\xi(2p)$  \\ \hline
Cr &$d^{3}$&10.777&6.755&--&--&--&0.035&--\\
&$d^{4}$&9.649&6.002&--&--&--&0.030&--\\
&$d^{5}$&8.357&5.146&--&--&--&0.025&--\\
&$d^{6}$&6.910&4.205&--&--&--&0.021&--\\
&$p^{5}d^{4}$&11.596&7.270&6.526&4.788&2.722&0.047&5.667\\
&$p^{5}d^{5}$&10.522&6.552&5.841&4.204&2.388&0.041&5.668\\
&$p^{5}d^{6}$&9.303&5.738&5.151&3.644&2.069&0.035&5.669\\
&$p^{5}d^{7}$&7.867&4.801&4.461&3.155&1.768&0.030&5.671\\
\end{tabular}
\end{ruledtabular}
\end{table*}
%%%%%%%%%%%%%%%%%%%%%%%%%%%%%%%%%%%%%%%%%%%%%%%%%%%%%%%%%%%%%%%
For Cr$_{8}$Te$_{9}$, we have assumed the nominal $d^{4}$ configuration and have employed two more charge-transfer states such as $d^{5}\underline{L}$ and $d^{6}\underline{L}^{2}$, where $\underline{L}$ denotes a hole in the Te $5p$ orbital.
Thus the initial state is expanded by a linear combination of $d^{4}$, $d^{5}\underline{L}$, $d^{6}\underline{L}^{2}$ and the final state is described by a linear combination of $\underline{2p}d^{5}$, $\underline{2p}d^{6}\underline{L}$, $\underline{2p}d^{7}\underline{L}^{2}$, where $\underline{2p}$ denotes a created hole in the $2p$ core level in the absorption final state.
Slater integrals and spin-orbit coupling constants for $d^{4}$, $d^{5}$ and $d^{6}$ configurations in the initial state and for $p^{5}d^{5}$, $p^{5}d^{6}$ and $p^{5}d^{7}$ in the XAS final states are listed in Table \ref{SI}.
To perform the CI calculation, four adjustable parameters are introduced as follows; the charge-transfer energy $\Delta\equiv E(d^{5}\underline{L})-E(d^{4})$, the Coulomb interaction energy $U_{dd}$ between the $3d$ electrons, the Coulomb energy $U_{cd}$ between the $2p$ core hole and $3d$ electrons, the hybridization energy $V_{\rm e_{g}} [=-\sqrt{3}(pd\sigma)]$ and the octahedral crystal field splitting $10Dq$.
We have neglected the hybridiation energy between the anion $p$ orbitals.

%*************************************
\begin{figure}
\includegraphics{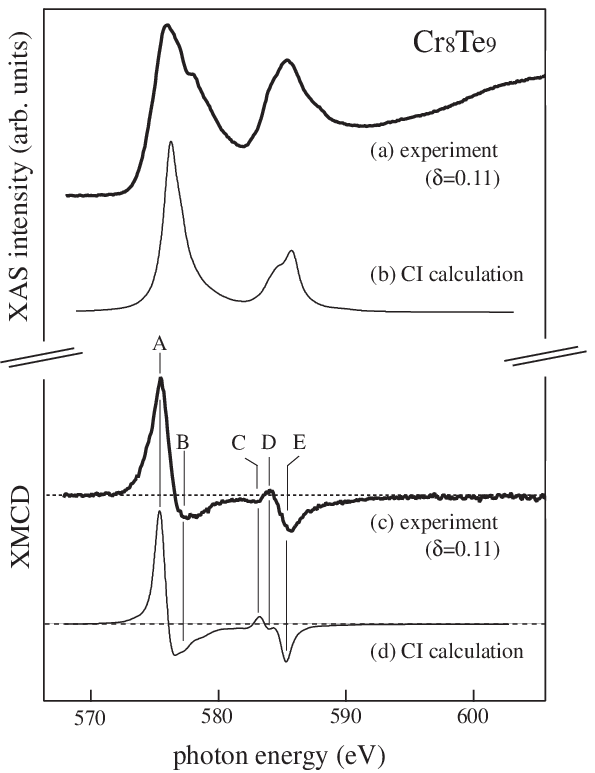}%
\caption{(a)(c) The experimental Cr $2p$ XAS and XMCD spectra of Cr$_{8}$Te$_{9}$ ({\it thin solid lines}).
(b)(d) The calculated XAS and XMCD spectra by the CI cluster model with full multiplets of [CrTe$_{6}$]$^{10-}$ cluster ({\it thick solid lines}).
}
\end{figure}
%*************************************

We have used the $U_{dd}\sim$2.3eV, which has been estimated from the Cr $M_{23}VV$ Auger-electron spectra and the self-convolution of the Cr $3d$-derived spectra.\cite{Shimada96}
We have also used the same value $V_{\rm e_{g}}$=1.3eV, which has been evaluated by using the formula $\displaystyle (pd\sigma)=\eta_{pd\sigma}\frac{\hbar^{2}}{m}\cdot\frac{r_{d}^{1.5}}{d^{3.5}}$ with $\eta_{pd\sigma}=-2.95$ and $r_{d}$(Cr)=-0.9\AA.\cite{Harrison}
Here, $U_{cd}$ has been fixed to $U_{dd}/U_{cd}=0.83$ and the relationship $(pd\sigma)/(pd\pi)=-2.0$ has been assumed.
Consequently, we have adjusted the charge transfer energy $\Delta$ and $10Dq$ to fit the experimental XAS and XMCD spectra of Cr$_{8}$Te$_{9}$.
Fig.4 (b) and (d) show the calculated XAS and XMCD spectra for Cr$_{8}$Te$_{9}$, which are broadened by the Lorenzian function with 2$\Gamma$=1eV.
It is noticed that the calculated XAS spectrum fits well with the experimental spectrum of Cr$_{8}$Te$_{9}$ including the shoulder structure in the $2p_{1/2}$ region except for the narrower band width in the calculated $2p_{3/2}$ XAS spectrum.
The calculated XMCD spectrum reproduces not only the dispersive XMCD feature at the $2p_{3/2}$ edge (A and B), but also the structures at $2p_{1/2}$ edge including the small positive structure found in the lower $h\nu$ region (C, D and E) as shown in Fig.4 (d).
It has been clarified that the hybridization effect between the Cr $3d$ and Te $5p$ states as well as the electron correlation effect of the Cr $3d$ electrons is quite important to understand the electronic states of Cr$_{1-\delta}$Te.
The calculated result also shows that the ground state wave function can be described as $\Psi_{g}=0.27|d^{4}\rangle+0.62|d^{5}\underline{L}\rangle+0.11|d^{6}\underline{L}^{2}\rangle$.
One notices from this result that the weight of the $d^{5}\underline{L}$ component is more than two times larger than that of the $d^{4}$ and the contribution of the $d^{6}\underline{L}^{2}$ is not negligible in the ground state, which has been derived from the strong hybridization between Cr $3d$ and Te $5p$ orbitals.
The averaged $3d$ electron number is 4.8, which is much larger than 4 (Cr$^{2+}$) as shown in Table \ref{CI}.
According to the CI calculation, the estimated $m_{\rm spin}$ and $m_{\rm orb}$ are 4.2$\mu_{\rm B}$ and -0.04$\mu_{\rm B}$, respectively.
Such a quite small $m_{\rm orb}$ stems from the dominant $d^{5}\underline{L}$ configuration in the ground state and is consistent with the estimated value with use of the sum rule as mentioned above.
The calculated $m_{\rm spin}$=4.2$\mu_{\rm B}$ is consistent with the formerly calculated result for the valence band photoemission spectrum of Cr$_{0.95}$Te, but is much larger than the values given by the band-structure calculation 3.3$\mu_{\rm B}$\cite{Dijikstra89} and the observed saturation magnetic moment $\sim$2.5$\mu_{\rm B}$.
In fact, we have neglected the Cr $d-d$ overlap along the $c$-axis in the CI calculation.
This may derive the larger spin magnetic moment in the calculation.
The observed discrepancy in the experimental XAS spectrum with the broader width accompanying the higher energy tail compared with the calculated one may also be ascribed to this lack of the $d-d$ interaction.

%*************************************
\begin{figure}
\includegraphics{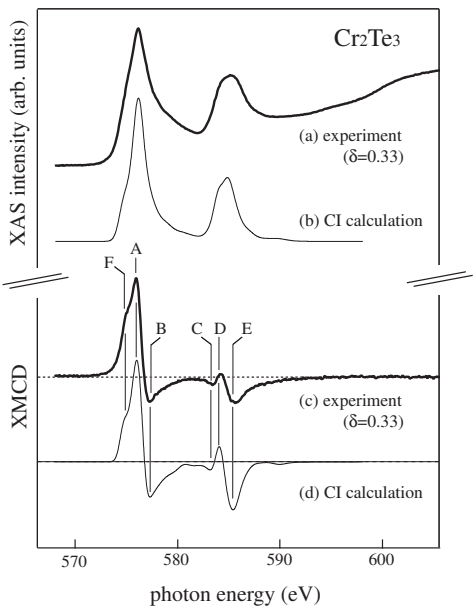}%
\caption{(a)(c) The experimental Cr $2p$ XAS and XMCD spectra of Cr$_{2}$Te$_{3}$ ({\it thin solid lines}).
(b)(d) The calculated XAS and XMCD spectra by the CI cluster model with full multiplets of [CrTe$_{6}$]$^{10-}$ cluster ({\it thick solid lines}.)
}
\end{figure}
%*************************************

%%%%%%%%%%%%%%%%%%%%%%%%%%%%%%%%%%%%%%%%%%%%%%%%%%%%%%%%%%%%%%
\begin{table}%[H] add [H] placement to break table across pages
\caption{
\label{CI}Parameters obtained from the analyses of the XAS and XMCD spectra of Cr$_{8}$Te$_{9}$ and Cr$_{2}$Te$_{3}$ with CI cluster model calculation (in units of eV).
We assumed $U_{dd}/U_{cd}=0.83$ and $(pd\sigma)/(pd\pi)=-2$.
}

\begin{ruledtabular}
\begin{tabular}{cccccccc}
&$\Delta$&$10Dq$&$U_{dd}$&$pd\sigma$&$n_{d}$&$m_{\rm spin}$&$m_{\rm orb}$\\ \hline
Cr$_{8}$Te$_{9}$&0.0&0.0&2.3&0.75&4.8&4.2&-0.04\\
Cr$_{2}$Te$_{3}$&-1.5&1.0&2.3&0.75&4.3&3.2&-0.03\\
\end{tabular}
\end{ruledtabular}
\end{table}
%%%%%%%%%%%%%%%%%%%%%%%%%%%%%%%%%%%%%%%%%%%%%%%%%%%%%%%%%%%%%%

%%%%%%%%%%%%%%Revised at 2002/12/29&30 (below)%%%%%%%%%%%%%%%
We have also applied the cluster model calculation to Cr$_{2}$Te$_{3}$, which has the nominal valency of 3+ (3$d^{3}$).
We have performed the calculation with the ground state wave function that is expanded by a linear combination of $d^{3}$, $d^{4}\underline{L}$, $d^{5}\underline{L}^{2}$ and $d^{6}\underline{L}^{3}$ configurations.
In this case, the charge transfer energy can be defined as $\Delta\equiv E(d^{4}\underline{L})-E(d^{3})$.
We have used the same values of $U_{dd}$=2.3eV, $U_{cd}=(U_{dd}/0.83)$ and $V_{\rm e_{g}}$=1.3eV as used for Cr$_{8}$Te$_{9}$.
Figs.5 (b) and (d) show the calculated XAS and XMCD spectra with the parameters $\Delta$=-1.5eV and $10Dq$=1.0eV.
We find a good correspondence between the calculated XAS/XMCD spectra and the experimental ones.
We recognize that the calculated XMCD spectrum not only reproduces the observed fundamental structures A-E but also the shoulder F as shown in Figs.5 (c) and (d) 
It is noted that we have obtained the negative value of $\Delta$ for Cr$_{2}$Te$_{3}$, which means that the ground state is not dominated by the $d^{3}$ configuration but by the $d^{4}\underline{L}$ and $d^{5}\underline{L}^{2}$ configurations because the energy differences $E(d^{4}\underline{L})-E(d^{3})$ and $E(d^{5}\underline{L}^{2})-E(d^{3})$ are expressed as $\Delta$(=-1.5eV) and $2\Delta+U_{dd}$(=-0.7eV), respectively when the hybridization ($V_{\rm e_{g}}$) is off.
That is, the most stable $d^{4}\underline{L}$ is formed by the charge transfer from the ligand Te $5p$ orbitals to the Cr $3d$ orbitals.
The resultant ground state wave function is $\Psi_{g}=0.11|d^{3}\rangle+0.48|d^{4}\underline{L}\rangle+0.36|d^{5}\underline{L^{2}}\rangle+0.05|d^{6}\underline{L}^{3}\rangle$.
From this result, the evaluated average electron number is found to be 4.3 and the calculated $m_{\rm spin}$ and $m_{\rm orb}$ are estimated to be 3.2$\mu_{\rm B}$ and -0.03$\mu_{\rm B}$, respectively as listed in Table II.
It is recognized that the small $m_{\rm orb}$ is consistent with the XMCD result, while the calculated spin magnetic moment is smaller than that for Cr$_{8}$Te$_{9}$ and is larger than the measured values of $\sim$2.0-2.7$\mu_{\rm B}$.
%%%%%%%%%%%%%%Revised at 2002/12/29 (above)%%%%%%%%%%%%%%%

Finally we discuss the $\delta$ dependence of the electronic structures in Cr$_{1-\delta}$Te.
We have already found that the XMCD spectral lineshapes are not changed so much with $\delta$ as shown in Fig.3.
This means that the change in the local Cr $3d$ electronic states is very weak when the Cr vacancy is introduced.
This interpretation can be supported by the CI cluster-model calculation.
As we have obtained in the CI cluster calculation with the parameters $\Delta=0.0eV$ and $U_{dd}=2.3eV$ for Cr$_{8}$Te$_{9}$, the relationship $\Delta<U_{dd}$ indicates that the Cr$_{8}$Te$_{9}$ can be classified in the charge transfer regime in the Zaanen-Zawatzky-Allen diagram.\cite{Zaanen85}
The early transition metal (Sc-Cr) compounds were originally classified in the Mott-Hubbard regime ($U_{dd}<\Delta$), but recent CI cluster calculation analyses of core-level photoemission spectra have shown that even the Cr oxides such as Cr$_{2}$O$_{3}$ and LaCrO$_{3}$ are reclassified in the intermediate region between the Mott-Hubbard and charge-transfer regimes ($U\simeq\Delta$).\cite{Bocquet96}
It is, therefore, considered that the classification of Cr tellurides in the charge-transfer regime is reasonable if we take into account the much smaller electron negativity of Te compared to oxygen.

%*************************************
\begin{figure}
\includegraphics{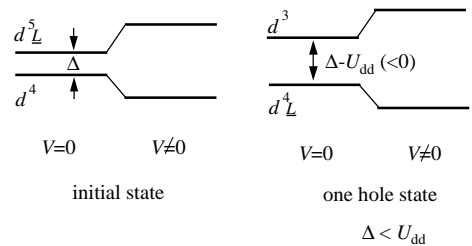}%
\caption{Schematic energy diagram of $d^{4}$ and $d^{5}\underline{L}$ initial states and $d^{3}$ and $d^{4}\underline{L}$ one hole (one electron removal) states.
For each state, the energy level with (without) the hybridization $V$ off (on) is shown at the left (right) hand side.
}
\end{figure}
%*************************************

Now we take into account the one hole (one electron removal) state that corresponds to the $3d$ photoemission final state.
To make the story as simple as possible, we describe the ground state wave function as $\Psi_{g}=\alpha|d^{4}\rangle+\beta|d^{5}\underline{L}\rangle$.
Then the one-hole state can be described as $\Psi_{f}=\alpha^{\prime}|d^{3}\rangle+\beta^{\prime}|d^{4}\underline{L}\rangle$.
Here the energy difference $E(d^{3})-E(d^{4}\underline{L}) $ can be represented as $\Delta-U_{dd}$.
According to the relationship $\Delta<U_{dd}$ as discussed above, the $d^{4}\underline{L}$ state is energetically lower compared to the $d^{3}$ state as shown in Fig.6.
In the other expression, if we introduce a hole in the [CrTe$_{6}$]$^{10-}$ cluster, the hole preferably stays in the Te $5p$ orbital.
The same interpretation can be applicable in the Cr deficient materials, Cr$_{1-\delta}$Te.
The created holes by the Cr vacancy in Cr$_{1-\delta}$Te are likely to be in the Te $5p$ orbitals, then the decrease of the $3d$ electron number with the increase of $\delta$ is weak, which is consistent with the calculated negative $\Delta$ value for Cr$_{2}$Te$_{3}$ and the present XMCD results.
This is also consistent with the calculated band structure of Cr$_{1-\delta}$Te, where the hole pocket derived from the Te $5p$ state appears around $\Gamma$ point in the Brillouin zone when the Cr vacancy is introduced in the material.\cite{Dijikstra89}

\section{Conclusion}
We have observed the Cr $2p$ XAS and XMCD spectra of Cr$_{8}$Te$_{9}$($\delta$=0.11), Cr$_{5}$Te$_{6}$($\delta$=0.17), Cr$_{3}$Te$_{4}$($\delta$=0.25) and Cr$_{2}$Te$_{3}$($\delta$=0.33).
The observed changes with the Cr vacancy in the experimental XMCD spectra of Cr$_{1-\delta}$Te are found to be quite small.
The experimental XAS and XMCD spectra of Cr$_{8}$Te$_{9}$ and Cr$_{2}$Te$_{3}$ have been compared with the result of the CI cluster-model calculation.
With the best fit parameters, $\Delta$=0eV, $U_{dd}$=2.3eV and $V_{\rm e_{g}}$=1.3eV, the calculated XAS and XMCD spectra have reproduced well the experimental spectra of Cr$_{8}$Te$_{9}$.
The obtained parameters show that Cr$_{1-\delta}$Te can be classified in the charge transfer type metal and the created holes in this compound should preferably exist in the ligand Te $5p$ orbitals, which is consistent with the obtained negative $\Delta$ for Cr$_{2}$Te$_{3}$ and the unchanged feature of the XMCD spectral lineshapes by the Cr defect concentration $\delta$.
These results are supported by the band structure calculation.

% If in two-column mode, this environment will change to single-column
% format so that long equations can be displayed. Use
% sparingly.
%\begin{widetext}
% put long equation here
%\end{widetext}

% Specify following sections are appendices. Use \appendix* if there
% only one appendix.
%\appendix
%\section{}

% If you have acknowledgments, this puts in the proper section head.
\begin{acknowledgments}
The authors would like to thank Dr. Y. Saitoh of the Japan Atomic Energy Research Institute for a fine adjustment of the monochromator and Professor A. Fujimori of the University of Tokyo and Professor T. Kanomata of Tohoku Gakuin University for their fruitful discussion.
This work was done under the approval of the SPring-8 Advisory Committee (Proposal No. 2000B0439-NS -np).
This work was supported by the Ministry of Education, Science, Sports and Culture.
\end{acknowledgments}

% Create the reference section using BibTeX:
%\bibliography{CrTe_XMCD}

%
\end{document}